\documentclass[pdflatex,sn-mathphys-num]{sn-jnl}

\usepackage{bm}
\usepackage{graphicx}%
\usepackage{multirow}%
\usepackage{amsmath,amssymb,amsfonts}%
\usepackage{amsthm}%
\usepackage{mathrsfs}%
\usepackage[title]{appendix}%
\usepackage{xcolor}%
\usepackage{textcomp}%
\usepackage{manyfoot}%
\usepackage{booktabs}%
\usepackage{algorithm}%
\usepackage{algorithmicx}%
\usepackage{algpseudocode}%
\usepackage{listings}%
\usepackage{siunitx}
\usepackage{dcolumn}
\usepackage[normalem]{ulem}
 
\def\u#1{_{\rm #1}}
\newcommand{\enquote}[1]{``#1''}
\newcommand{\ket}[1]{| #1 \rangle}
\newcommand{\bra}[1]{\langle #1 |}
\newcommand{\ketbra}[2]{| #1 \rangle \langle #2 |}
\newcommand{\expect}[1]{\langle #1 \rangle}

\def\g2{g^{(2)}}

\DeclareRobustCommand{\erase}{\bgroup\markoverwith{\textcolour{red}{\rule[.5ex]{2pt}{0.4pt}}}\ULon}

\begin{document}

\title[Article Title]{Experimental zero-added-loss multiplexing
Bell-pair source for long-haul quantum networks}


\author*[1]{\fnm{Yoshiaki} \sur{Tsujimoto}}\email{tsujimoto@nict.go.jp}

\author[1,2]{\fnm{Daiki} \sur{Ichii}}

\author[3,4]{\fnm{Rikizo} \sur{Ikuta}}

\author[1]{\fnm{Mikio} \sur{Fujiwara}}

\author[1,2]{\fnm{Masahiro} \sur{Takeoka}}

\author[1]{\fnm{Go} \sur{Kato}}

\author[1]{\fnm{Kentaro} \sur{Wakui}}

\affil[1]{\orgname{Advanced ICT Research Institute, National Institute of Information and Communications Technology~(NICT)}, \city{Koganei}, \state{Tokyo} \postcode{184-8795}, \country{Japan}}

\affil[2]{\orgname{Department of Electronics and Electrical Engineering, Keio University}, \city{Yokohama}, \state{Kanagawa} \postcode{223-8522}, \country{Japan}}

\affil[3]{\orgname{Graduate School of Engineering Science, The University of Osaka}, \city{Toyonaka}, \state{Osaka} \postcode{560-8531}, \country{Japan}}

\affil[4]{\orgname{Center for Quantum Information and Quantum Biology, The University of Osaka}, \city{Toyonaka}, \state{Osaka} \postcode{560-0043}, \country{Japan}}


\abstract{
Boosting the communication rate of quantum networks is a central challenge in quantum information science. Recently, an efficient entanglement distribution scheme employing quasi-deterministic Bell-pair sources based on time-frequency multiplexing, referred to as zero-added-loss multiplexing~(ZALM), has been proposed. Its implementation, however, requires high-fidelity entanglement swapping across densely multiplexed time-frequency modes, which has remained an experimental challenge.
Here we demonstrate entanglement swapping across 16 parallel frequency modes with a high average fidelity of 93.9$\pm$\SI{1.4}{\%}. Notably, polarization-entangled photon pairs in each frequency mode are spectrally single-mode using only off-the-shelf 50-GHz dense wavelength-division multiplexing~(DWDM) filters, eliminating the need for additional narrowband filtering.
Furthermore, in order to fully exploit the temporal degree of freedom, the pump pulse is operated with a repetition frequency of \SI{3.0}{GHz}. By combining the frequency and time multiplexing, the total swapping rate reaches 5.38$\pm$0.17\,\si{pairs\,s^{-1}}, which corresponds to the ZALM Bell-pair rate of \SI{8.2e2}{pairs\,s^{-1}}. Our results establish the key experimental capabilities required for ZALM and demonstrate a scalable route toward practical high-rate quantum repeaters and long-haul quantum networks.
}

\maketitle

\section*{INTRODUCTION}\label{sec1}
The realization of quantum networks is a central goal of quantum information science, with applications ranging from secure communication~\cite{Pirandola:20} to distributed quantum computation~\cite{PhysRevA.89.022317} and sensing~\cite{Zhang_2021, Komar2014}. In order to achieve both high rates and high fidelity, one promising route is to exploit multiple modes in various photonic degrees of freedom, including frequency~\cite{Aktas2016,Wengerowsky2018,PhysRevLett.134.230801,Mueller:24}, spatial modes~\cite{Ortega:24}, and orbital angular momentum~\cite{PhysRevLett.127.093601,Liu2020} for multiplexed quantum communication. In particular, frequency multiplexing is highly attractive due to its compatibility with existing fiber-based telecommunication infrastructure based on dense wavelength division multiplexing~(DWDM) technology and its ability to harness the intrinsically broad spectral bandwidth of photon pairs generated via spontaneous parametric down-conversion~(SPDC)~\cite{Aktas2016,Wengerowsky2018,PhysRevLett.134.230801,Mueller:24,PhysRevLett.123.193603,Wakui:20}.

\begin{figure}[t]
 \begin{center}
  \includegraphics[width=\columnwidth]{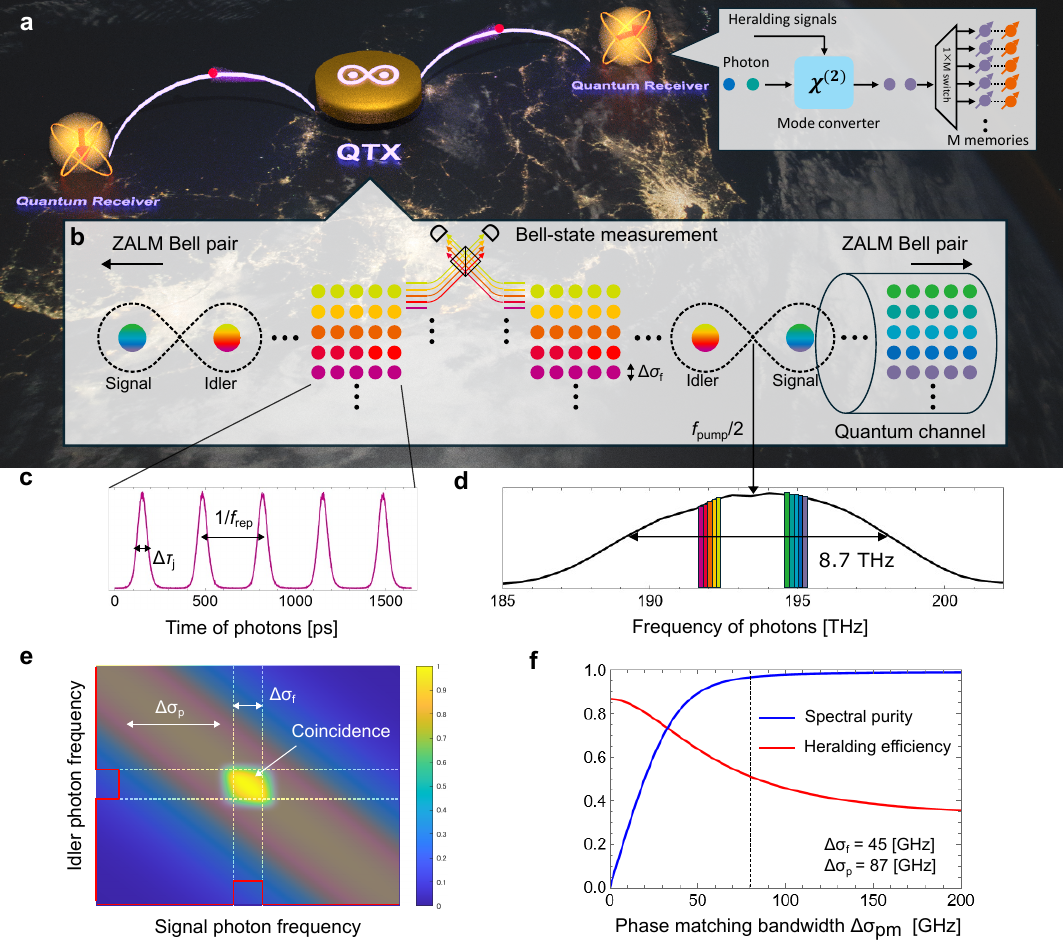}
  \caption{{\bf ZALM architecture based on time-frequency-multiplexed entanglement swapping.} {\bf a}~Concept of the entanglement distribution using ZALM. Background image adapted from an image courtesy of the Earth Science and Remote Sensing Unit, NASA Johnson Space Center~(ISS053-E-209380). {\bf b}~Implementation of the QTX using time-frequency-multiplexed entanglement swapping based on SPDC. {\bf c}~The temporal distribution of the photon counts. The repetition frequency and the timing jitter of the photon detector are $f\u{rep}=$ \SI{3.0}{GHz} and $\Delta\tau\u{j}=$ \SI{39}{ps}, respectively.
   {\bf d}~The spectrum of the SPDC photons. The center frequency is chosen to be $f\u{pump}/2=$ \SI{193.4}{THz}. {\bf e}~The joint spectral amplitude of the photon pair in a single DWDM channel. We focus on a pair of output modes of the DWDM filters in the signal and idler arms. {\bf f}~The relation among the phase-matching bandwidth, spectral purity and heralding efficiency of the swapped photon pair under the condition of $\Delta\sigma\u{f}=$ \SI{45}{GHz} and $\Delta\sigma\u{p}=$ \SI{87}{GHz}. 
  The black dotted line shows the experimental phase-matching bandwidth of $\Delta\sigma\u{pm}=$ \SI{80}{GHz}.
  \label{fig1:Concept}} 
 \end{center}
\end{figure}

Recently, an efficient entanglement distribution scheme was proposed that exploits frequency multiplexing using the broad spectral bandwidth of the SPDC photon pairs, which is called zero-added-loss multiplexing~(ZALM)~\cite{PhysRevApplied.19.054029}. 
In the ZALM architecture, a quantum transmitter~(QTX) is placed at the central node, and Bell pairs are distributed to neighboring nodes equipped with quantum mode converters and quantum memories, as shown in Fig.~\ref{fig1:Concept}a. 
This configuration is known as a midpoint-source configuration and enables a high-rate entanglement distribution with a limited number of quantum memories compared to other architectures, such as the meet-in-the-middle configuration~\cite{Jones_2016}. The QTX employs a ZALM Bell-pair source based on local frequency-multiplexed entanglement swapping~\cite{PhysRevLett.71.4287} to achieve quasi-deterministic entanglement generation, thereby circumventing the switching losses that have rendered previous multiplexing schemes inefficient~\cite{PhysRevA.84.052326,PhysRevApplied.17.034071}.
Using classical information from the QTX identifying the frequency mode in which the heralding event occurred, the generated photon pairs are converted into the desired time-frequency mode by mode converters at the quantum receivers.

A key advantage of ZALM is that it leverages not only frequency multiplexing but also the ultrafast photon-pair generation capability of SPDC~\cite{Wakui:20}, enabling Bell-pair generation with both high probability and high repetition rate.
In fact, the ZALM proposal~\cite{PhysRevApplied.19.054029} typically requires a DWDM channel bandwidth of \SI{12.5}{GHz}, a SPDC spectral bandwidth of \SI{10}{THz}, and a pump repetition rate of \SI{30}{GHz} to realize dense time-frequency multiplexing.
Previous experimental demonstrations of frequency multiplexing have been limited to the direct distribution of entangled photon pairs~\cite{Aktas2016,Wengerowsky2018,PhysRevLett.134.230801,Mueller:24}.
Frequency multiplexing has also been explored for entanglement swapping between time-frequency entangled photons~\cite{PhysRevLett.128.063602} and quantum fusion of independent quantum networks~\cite{Huang2026}.
However, the wavelengths~\cite{PhysRevLett.128.063602} and architectures~\cite{Huang2026} demonstrated in these studies, combined with their repetition rates in the tens of MHz range, made it difficult to apply them to the ZALM architecture, particularly with its goal of increasing the entanglement-swapping rate through time-frequency multiplexing.

Here, we demonstrate the ZALM Bell-pair source through high-fidelity entanglement swapping across 16 parallel frequency modes in the telecom C-band for long-haul communication, at a GHz-level repetition rate.
To enable direct frequency multiplexing with off-the-shelf 50-GHz DWDM filters, we combine pump-pulse waveshaping with careful selection of the SPDC crystal length
such that each frequency channel is spectrally single-mode, leading to an average Hong-Ou-Mandel~(HOM) interference~\cite{PhysRevLett.59.2044} visibility of $95.0\pm$\SI{0.7}{\%} across all 16 modes. 
In addition to frequency multiplexing, we fully exploit the temporal degree of freedom by operating the pump laser at a repetition rate of \SI{3.0}{GHz}~\cite{Wakui:20,Tsujimoto:21}. This record-high repetition rate allows us to increase the generation rate of the SPDC photon pairs without compromising the interference visibility of the photons. By jointly exploiting the frequency and temporal degrees of freedom, we achieve a total swapping rate of 5.38$\pm$0.17 pairs $\text{s}^{-1}$ and an average entanglement fidelity of $93.9\pm$\SI{1.4}{\%}, which corresponds to the ZALM Bell-pair rate of 8.2$\times\text{10}^\text{2}$ pairs $\text{s}^{-1}$. These results establish a new performance regime for photonic entanglement swapping beyond those accessed in previous experiments. 
Moreover, simulations based on experimental parameters show that, by fully exploiting the 8.7-THz SPDC spectrum, the ZALM Bell-pair rate can reach $10^5$ pairs $\text{s}^{-1}$ while maintaining an entanglement fidelity of \SI{91}{\%}. 
Our results establish dense time-frequency-multiplexed entanglement swapping as a viable route toward high-rate quantum communication and constitute a key enabling step toward the ZALM architecture for long-haul quantum networks.

\section*{RESULTS}\label{sec2}
\subsection*{Experimental challenges for ZALM Bell-pair source}
\label{subsubsec1}
In this work, we focus on the QTX shown in Fig.~\ref{fig1:Concept}b and address the key experimental challenges in realizing ZALM Bell-pair source.
At each SPDC source, the lower-frequency~(idler) and higher-frequency~(signal) components of the SPDC photon pair are spatially separated, and the pairs that satisfy energy conservation are entangled in polarization degree of freedom.
The idler photons are interfered on a 50:50 beamsplitter~(BS) to perform a time-frequency-resolved Bell-state measurement~(BSM), thereby allowing time-frequency-multiplexed entanglement swapping. Under the condition that coincidence events between idler photons of the same frequency are detected at the BSM node, a quasi-deterministic generation of entangled photon pairs becomes possible when the number of frequency modes is sufficiently large~\cite{PhysRevApplied.19.054029}.
A key challenge in QTX is to maximize the number of modes in the available time-frequency space. However, compared with simple transmission of entangled photon pairs, entanglement swapping imposes more stringent constraints on time-frequency-mode multiplexing. 

For the temporal degree of freedom, the number of temporal modes is increased by increasing the pump repetition rate $f\u{rep}$ as shown in Fig.~\ref{fig1:Concept}c. However, when the pulse spacing becomes shorter than the timing jitter of the photon detectors $\Delta\tau\u{j}$, photons from neighboring pulses can enter the coincidence detection window. 
We emphasize that this effect is much more pronounced than in the case of a direct distribution of entangled photon pairs. In the direct distribution, even under the conditions described above, the main consequence is merely a small admixture of error events due to multiphoton generation~\cite{Wakui:20}. In contrast, in entanglement swapping, insufficient temporal resolution of the pulse train directly results in an increase in the number of modes contributing to the HOM interference, which dramatically degrades the interference visibility~\cite{Tsujimoto:21}. Thus, a time-resolved detection of the pulse train satisfying the condition $1/f\u{rep}\gg\Delta\tau\u{j}$ is necessary. 

For frequency multiplexing, the accessible bandwidth is limited by the phase-matching bandwidth of the nonlinear optical crystal for SPDC, within which a higher multiplexing density enables the utilization of a larger number of frequency modes as shown in Fig.~\ref{fig1:Concept}d. Here, an important trade-off exists between the spectral bandwidth of the pump pulses and that of each multiplexed mode. For each frequency mode that can be separated in parallel by the DWDM filters, spectral purity is guaranteed as long as the spectral bandwidth of the pump pulses $\Delta\sigma\u{p}$ is sufficiently larger than the filter bandwidth $\Delta\sigma\u{f}$~\cite{zukowski1995,rarity1995} as shown in Fig.~\ref{fig1:Concept}e. However, if $\Delta\sigma\u{p}$ is excessively larger than $\Delta\sigma\u{f}$, the following two detrimental effects occur.  
The first issue is inter-channel cross
talk of DWDM filters. Unlike the case of continuous-wave pumping, where strong frequency-correlation is imposed by energy conservation, the finite bandwidth of the pump pulses allows photon pairs to leak into adjacent frequency modes.
The second issue is a photon loss. When the spectrally pure mode is further subdivided, it effectively results in a photon loss even if the filter transmittance is \SI{100}{\%}~\cite{PhysRevA.95.061803}. Consequently, it is necessary to tune the pump bandwidth so as to achieve the required spectral purity while minimizing the photon loss as well as the inter-channel crosstalk.

In this work, we employ type-0 periodically poled lithium niobate waveguides~(PPLN/Ws) for the SPDC, which provide a broad phase-matching bandwidth. The experimentally measured spectrum of the generated photon pairs is shown in Fig.~\ref{fig1:Concept}d, exhibiting a bandwidth of \SI{8.7}{THz} FWHM. Using standard 50-GHz DWDM filters, we implement frequency multiplexing over 16 frequency modes in each of the signal and idler modes. 
It should be noted that these 16 DWDM output channels merely conform to the standard specifications of a single commercially available device, and it is easy to increase the number of channels further.
The experimental parameters are optimized to satisfy the constraints as follows. First, the SPDC sources were operated at a repetition frequency of $f\u{rep}=$ \SI{3.0}{GHz}, which remains fully resolvable with our superconducting single-photon detectors~(SSPDs) with an average timing jitter of $\Delta\tau\u{j}=$ \SI{39}{ps} as shown in Fig.~\ref{fig1:Concept}c.
Second, the joint spectral amplitude~(JSA) was engineered to simultaneously achieve high spectral purity and high heralding efficiency.
In our experiment, not only $\Delta\sigma\u{p}$ but also the phase-matching bandwidth $\Delta\sigma\u{pm}$ of the PPLN/W for SPDC contributes to the trade-off between spectral purity and photon transmittance. Thus, after maximizing $\Delta\sigma\u{p}$, we selected an appropriate PPLN/W length corresponding to the desired $\Delta\sigma\u{pm}$.
Fig.~\ref{fig1:Concept}f shows the relationship between the spectral purity and the heralding efficiency~(transmittance) of the photon pair as a function of $\Delta\sigma\u{pm}$, while keeping $\Delta\sigma\u{f}$ and $\Delta\sigma\u{p}$ fixed~(see Methods for details). 
The simulation shows that $\Delta\sigma\u{pm}$ corresponding to a \SI{2.0}{cm}-long PPLN/W simultaneously provides a spectral purity of \SI{97}{\%} and a heralding efficiency of \SI{57}{\%}.

\subsection*{Experimental setup}
\label{subsubsec2}
\begin{figure}[t]
 \begin{center}
  \includegraphics[width=\columnwidth]{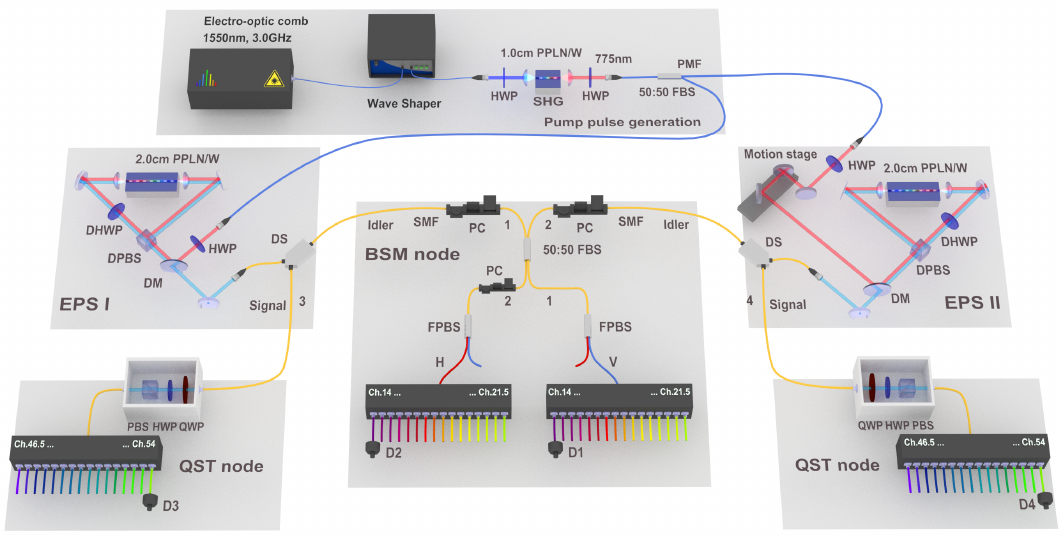}
  \caption{{\bf Experimental setup for time-frequency-multiplexed entanglement swapping.} 
  Pump pulses centered at \SI{775}{nm} with a 3.0-GHz repetition rate is prepared by 
  second harmonic generation~(SHG) of the electro-optic comb centered at \SI{1550}{nm}~(\SI{193.4}{THz}).   The SHG pulses are used to pump entangled photon pair sources~(EPSs) I and I\!I based on 
  PPLN/Ws. The signal and idler photons are separated into the shorter and longer wavelengths with respect to \SI{1550}{nm} by dichroic separators~(DSs) integrated in the fiber pigtail. The signal and idler photons are sent to the quantum state tomography~(QST) nodes and Bell state measurement~(BSM) node, respectively. After 16 frequency modes in each of signal and idler photons are extracted, photons are detected by SSPDs~(D1-D4). The four-fold coincidence counts are recorded by time-to-digital converter~(not shown). The details are given in main text.
 HWP: half waveplate, QWP: quarter waveplate, PBS: polarizing beamsplitter, PMF: polarization maintaining fiber, SMF: single-mode fiber, DHWP: dual-wavelength HWP, DPBS: dual-wavelength PBS, DM: dichroic mirror, PC: polarization controller, 50:50 FBS: 50:50 fiber-based beamsplitter, FPBS: fiber-based PBS.
  \label{fig2:Experiment}} 
 \end{center}
\end{figure}

The experimental setup for time-frequency-multiplexed entanglement swapping is shown in Fig.~\ref{fig2:Experiment}. 
We prepare pump pulses with $f\u{rep}=$\SI{3.0}{GHz} by doubling the frequency of the fundamental telecom pulses via second harmonic generation~(SHG) using a 1.0-cm-long type-0 PPLN/W. The center frequency of the fundamental pulse is chosen as \SI{193.4}{THz}, which corresponds to Ch.34 of the ITU channel. 
Using a wave shaper, we optimized the bandwidth and dispersion of the fundamental pulse to fully utilize the phase-matching bandwidth of the 1.0-cm-long PPLN/W while minimizing the pulse duration of the SHG pulse.
The FWHMs of the SHG pulse in the frequency~(time) domain were measured to be \SI{87}{GHz}~(\SI{5.0}{ps}), respectively, assuming $\mathrm{sech^2}$ spectral~(temporal) profiles.
The time-bandwidth product is calculated as 0.44, which is larger than the transform-limited value of 0.315, indicating a slight deviation from the ideal $\mathrm{sech^2}$ spectral and temporal profiles.
The SHG pulses are used to pump entangled photon pair sources~(EPSs) I and I\!I. 
Each EPS consists of a 2.0-cm-long type-0 PPLN/W in a Sagnac interferometer with a dual-wavelength polarizing beamsplitter~(DPBS). 
The phase-matching function of the 2.0-cm-long PPLN/W is well fitted by a super-Gaussian with order 1.4 and a FWHM of $\Delta\sigma_{\mathrm{pm}}=80~\mathrm{GHz}$, which is slightly narrower than the pump bandwidth $\Delta\sigma_p$~(see Methods for the fitting function). 

The signal and idler wavelengths are spatially separated by fiber-based dichroic separators~(DSs). The idler photons are sent to the BSM node to perform the time-frequency-resolved BSM. Polarization rotations during propagating single-mode fibers are compensated by using in-line polarization controllers.
After mixing two idler photons by a fiber-based 50:50 beamsplitter, horizontally~(H) and vertically~(V) polarized components in left and right output modes are respectively extracted using fiber-based PBSs. Then, 16 frequency modes are extracted using 50-GHz DWDM filters~(from ITU Ch.14 to Ch.21.5) followed by photon detection by SSPDs~(D1 and D2). The measured transmission spectrum of the flat-top DWDM filters is well fitted by a super-Gaussian with order 2.5 and a FWHM of $\Delta\sigma\u{f}=$\SI{45}{GHz}. The average timing jitter per channel is \SI{39}{ps}, including the jitter of a time-to-digital converter~(TDC), which is much smaller than the pulse interval of $1/f\u{rep}=$\SI{333}{ps} allowing time-resolved measurements.
The signal photons are sent to the quantum state tomography~(QST) nodes to perform polarization measurements on the swapped state. 
The polarization rotation in each arm is canceled by rotating a pair of a quarter waveplate and half waveplate~(not shown). After performing polarization projections, corresponding 16 frequency modes are extracted using \SI{50}{GHz} DWDM filters~(from ITU Ch.46.5 to Ch.54) followed by photon detection by SSPDs~(D3 and D4). 

\subsection*{Frequency-multiplexed HOM interference}
\label{subsubsec3}
\begin{figure}[t]
 \begin{center}
  \includegraphics[width=\columnwidth]{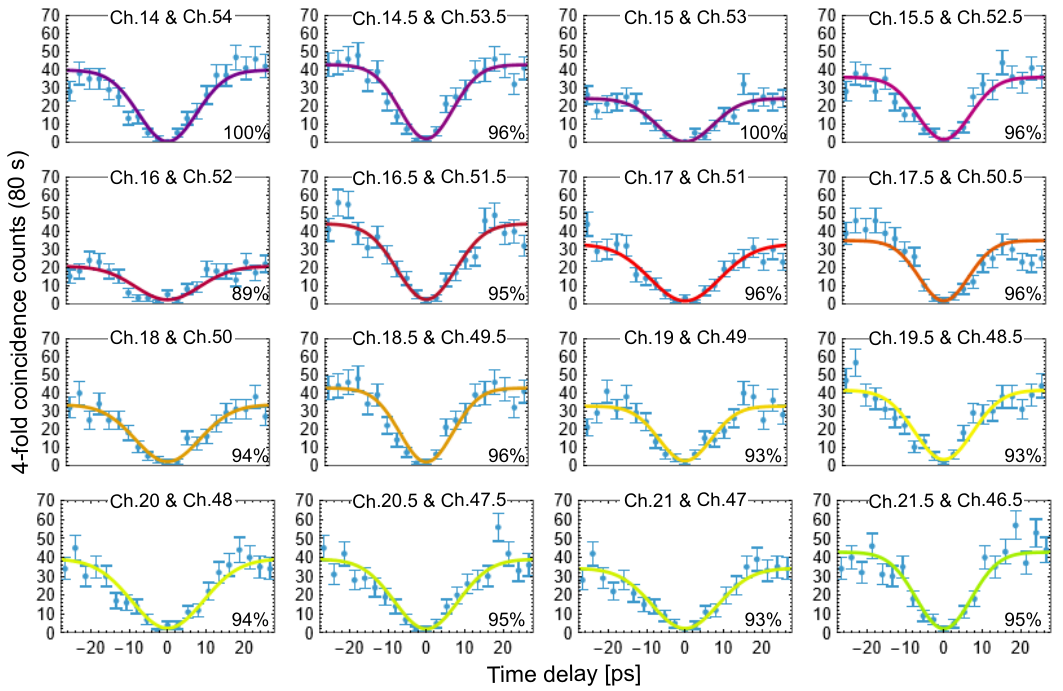}
  \caption{{\bf Frequency-multiplexed HOM dips.} The HOM dips and visibilities for 16 frequency modes in parallel. Each frequency-mode pair is denoted as Ch.$n\u{i}$ \& Ch.$n\u{s}$, where Ch.$n\u{i}$~(Ch.$n\u{s}$) represents the ITU channel of the idler~(signal) photon, respectively.  
  The measurement time was \SI{80}{s} for each point. 
  The error bars correspond to the standard deviations with the assumption of Poisson statistics for the photon counts. The solid curves are the maximum-likelihood fits using a Gaussian dip model assuming Poisson counting statistics.
  \label{fig3:HOM}} 
 \end{center}
\end{figure}

To assess the indistinguishability of the heralded idler photons in each ITU channel, the HOM interference experiment was performed. 
In this experiment, the four-fold coincidence events of H-polarized components were recorded by a TDC with changing the relative delay of two idler photons using a motion stage placed at EPS I\!I in Fig.~\ref{fig2:Experiment}. 
In postprocessing, we apply coincidence windows to the detection signals from D2, D3, and D4, conditioned on a detection event at D1. 
The width of each coincidence window, $\tau\u{w}$, was chosen to be \SI{100}{ps} such that
$\Delta\tau\u{d} < \tau\u{w} < 1/f\u{rep}$,
where $\Delta\tau\u{d}$ is the temporal width of the coincidence peak given by
$\Delta\tau\u{d}=\sqrt{2\Delta\tau\u{j}^2+T\u{c}^2}=\SI{57}{ps}$,
with $\Delta\tau\u{j}=$\SI{39}{ps} being the detector timing jitter and
$T\u{c}=\SI{15}{ps}$ the photon coherence time.
The repetition period was $1/f\u{rep}=\SI{333}{ps}$. The observed HOM dips for 16 ITU channel combinations~(from ITU Ch.14 \& Ch.54 to Ch.21.5 \& Ch.46.5) are shown in Fig.~\ref{fig3:HOM}. 
The HOM visibilities were obtained by fitting the measured coincidence counts to the model using maximum-likelihood estimation assuming Poisson counting statistics. The average visibility is $95.0\pm$\SI{0.7}{\%}, indicating that the heralded idler photons in each mode present a nearly perfect indistinguishability.
The experimentally obtained values can be evaluated using the following formula~\cite{Tsujimoto:21,PhysRevApplied.19.014008}
\begin{equation}
    V\u{th}=\frac{P}{1+\frac{\zeta g^{(2)}\u{I}+\zeta^{-1}g^{(2)}\u{I\!I}}{2}}, 
    \label{eq:Vth}
\end{equation}
where $P$ is the spectral purity of the photon pair, assuming identical Schmidt-mode structures for the two input photons, $\zeta$ is the intensity ratio of the heralded idler photons from EPS~I and EPS~I\!I, and $g_{\mathrm{I}}^{(2)}$ and $g_{\mathrm{I\!I}}^{(2)}$ are the corresponding intensity correlation functions.
Using the experimental values of ITU Ch.21.5 \& Ch.46.5 as $\zeta=1.0$, $g\u{I}^{(2)}=2.8\times10^{-2}$ and $g\u{I\!I}^{(2)}=3.5\times10^{-2}$,
and $P=$\SI{97}{\%} estimated by the simulation, $V\u{th}$ is calculated as \SI{94}{\%}, which is in reasonable agreement with the experimental value.

\subsection*{Generation of ZALM Bell-pairs}
\label{subsubsec4}

\begin{figure}[t]
 \begin{center}
  \includegraphics[width=\columnwidth]{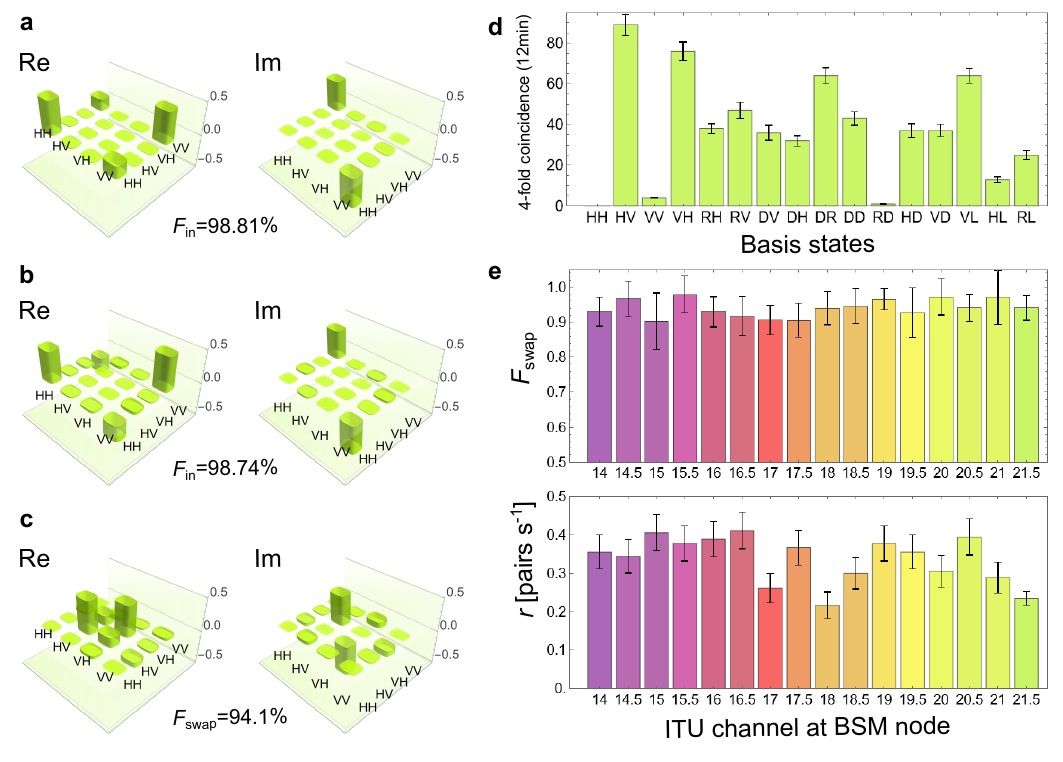}
  \caption{{\bf Results for time-frequency-multiplexed entanglement swapping} 
  {\bf a,b}~The reconstructed density matrices of input states. {\bf c}~The reconstructed density matrix of the swapped state heralded by the BSM in ITU Ch.21.5. {\bf d}~The four-fold coincidence counts for 16 measurement basis states. {\bf e}~The fidelities of the swapped states~($F\u{swap}$) and the four-fold coincidence count rates~($r$) for 16 independent ITU channel combinations. The integration time was \SI{3}{min} per basis state, except for ITU Ch.~21.5~(\SI{12}{min} per basis state).
  \label{fig4:swapping}} 
 \end{center}
\end{figure}

We first characterize the initial quantum states generated by EPS~I and EPS~I\!I. As a specific example, we present experimental results for the case where the idler and signal photon frequencies correspond to ITU Ch.21.5 \& Ch.46.5, respectively. The average photon numbers of the photon pairs are estimated to be $3.3\times10^{-3}$ for EPS~I and $4.6\times10^{-3}$ for EPS~I\!I, respectively. The density operators of the initial states $\rho_{13}$ and $\rho_{24}$, generated by EPS~I and EPS~I\!I, respectively, are reconstructed by performing QST as shown in Figs.~\ref{fig4:swapping}a and b, respectively. 
In our setup, since the relative phases between the H and V polarizations in the initial states are not adjusted, the initial states correspond to the Bell state $\ket{\Phi^+}:=(\ket{HH}+\ket{VV})/\sqrt{2}$ with its phase rotated by a local unitary operation as $R(\phi)\ket{\Phi^+}$, where $\ket{H(V)}$ represents the horizontally~(vertically) polarized single-photon state and $R(\phi):=\ketbra{H}{H}+e^{i\phi}\ketbra{V}{V}$.
The fidelity to the maximally entangled 
state is defined by $F\u{in}:=\mathrm{max}_{\phi}\bra{\Phi^+}R^\dagger(\phi)\rho R(\phi)\ket{\Phi^+}$, which are estimated to be $98.81\pm$\SI{0.03}{\%} and $98.74\pm$\SI{0.03}{\%} for $\rho_{13}$ and $\rho_{24}$, respectively. 
These values indicate that the nearly ideal initial states are prepared. 

The entanglement swapping experiments were performed in parallel across 16 frequency modes.
We briefly describe the entanglement-swapping procedure for each frequency mode in the present experiment. 
When a BSM corresponding to a projection onto $\ket{\Psi^-}:=(\ket{HV}-\ket{V\!H})/\sqrt{2}$ is performed on the idler photons in modes 1 and 2, the unnormalized state after a successful BSM is given by 
\begin{eqnarray}
_{12}\bra{\Psi^-}R_3(\phi_3)\ket{\Phi^+}_{13}\otimes R_4(\phi_4)\ket{\Phi^+}_{24}
&=&\frac{1}{2}R_3(\phi_3)R^\dagger_3(\phi_4)\ket{\Psi^-}_{34}, 
\end{eqnarray}
which is a state obtained from $\ket{\Psi^-}$ by a local unitary phase rotation by an angle $\phi_3-\phi_4$. 
In our experiment, the success probability of the BSM is $P\u{BSM}=1/8$ since only $\ket{V\!H}_{12}$ is measured at the BSM node. 
In the experiment, QST is performed on the signal photons in modes 3 and 4, conditioned on coincidence detection between D1 and D2. As in the HOM experiment, we set the coincidence windows with $\tau\u{w}=$\SI{100}{ps}. The reconstructed density operator of the swapped state $\rho_{34}$ and the four-fold coincidence counts for each measurement basis state are shown in Fig.~\ref{fig4:swapping}c and d, respectively. The fidelity is
$F_{\mathrm{swap}}:=\max_{\phi}\bra{\Psi^-}R^\dagger(\phi)\rho_{34}R(\phi)\ket{\Psi^-}
=94.1\pm\SI{3.5}{\%}$,
indicating that strong entanglement is generated in the swapped state.
Maximization over $\phi$ was performed independently for each frequency mode.
The swapping rate $r$ is calculated as $\sum_{i,j\in\{H,V\}}C_{ij}/T_{\mathrm{meas}}$, where $C_{ij}$ denotes the four-fold coincidence count for the polarization basis state $\ket{ij}$ and $T_{\mathrm{meas}}$ is the integration time, yielding $r=0.27$ pairs~s$^{-1}$. 
The measured values of $F_{\mathrm{swap}}$ and $r$ for the 16 ITU channel combinations are summarized in Fig.~\ref{fig4:swapping}e.
The rate-weighted average fidelity of the swapped Bell pairs
$F_{\mathrm{exp}}
:=
\sum_k r_kF_{\mathrm{swap}}^{(k)}/\sum_k r_k$,
and the total swapping rate,
$r_{\mathrm{tot}}
:=
\sum_k r_k,
$
are measured to be $93.9\pm$\SI{1.4}{\%} and $5.38\pm0.17$ pairs~s$^{-1}$, respectively. Here, $r_k$ and $F_{\mathrm{swap}}^{(k)}$ denote the swapping rate and the entanglement fidelity for the $k$th ITU channel combination, respectively.
We estimate the ZALM Bell-pair rate $r_{\mathrm{ZALM}}$ by correcting only for the detection losses in signal modes 3 and 4, defined as 
\begin{equation}
r_{\mathrm{ZALM}}:=\sum_{i,j\in\{H,V\}}\frac{C_{ij}}{\eta_{i_3}\eta_{j_4}T_{\mathrm{meas}}},
\end{equation}
where $\eta_{H_3}$, $\eta_{V_3}$, $\eta_{H_4}$, and $\eta_{V_4}$ are \SI{7.4}{\%}, \SI{7.4}{\%}, \SI{8.0}{\%}, and \SI{8.9}{\%}, respectively, yielding $r_{\mathrm{ZALM}}=8.2\times10^{2}$ pairs~s$^{-1}$. The intrinsic heralding efficiencies of \SI{57}{\%} due to spectral filtering in each signal modes 3 and 4 are not corrected for, as this filtering is required to achieve the measured spectral purity~(see Methods for details).

\section*{DISCUSSION}
\label{sec3}

\begin{figure}[t]
 \begin{center}
  \includegraphics[width=\columnwidth]{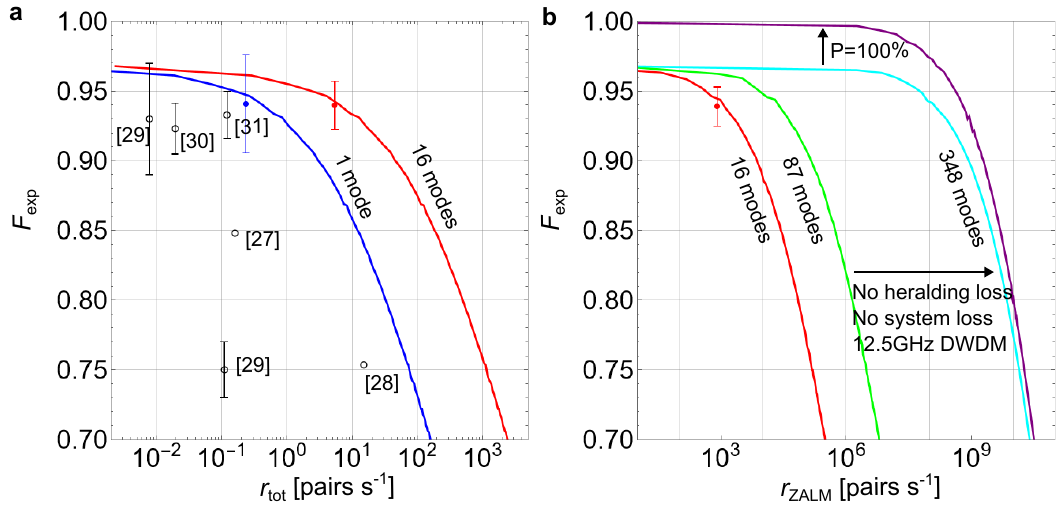}
  \caption{{\bf Simulation results for the rate-fidelity trade-off} 
  {\bf a}~The relationship between $r\u{tot}$ and $F\u{exp}$. The curves are obtained by the theoretical simulation using experimental parameters. The filled and hollow circles corresponds to our experimental results and previous results~\cite{Wu_2013,Jin2015,Tsujimoto2018,Samara_2021,wang2026}, respectively. 
  {\bf b}~The relationship between $r\u{ZALM}$ and $F\u{exp}$. The red dot and curve show our experimental result and simulation result with current experimental parameters, respectively. The green curve corresponds to the case of $P\u{BSM}=1/2$ with frequency multiplexing over the full SPDC bandwidth. The cyan curve shows the simulated performance with improved optical loss and frequency-multiplexing density, while the purple curve further includes an improvement in spectral purity.
  \label{fig5:ZALMrate}} 
 \end{center}
\end{figure}

In this section, we clarify the trade-off between $r\u{tot}$ and $F\u{exp}$ using a theoretical model together with experimental parameters. We further discuss future prospects for increasing $r\u{ZALM}$ by improving the key experimental parameters and estimate the achievable performance. We modeled the experimental system using the Gaussian-state characteristic-function method~\cite{Takeoka_2015,PhysRevA.93.042328,Tsujimoto_2020} and performed numerical simulations using the independently characterized system parameters, including optical loss, spectral purity, and detector quantum efficiency~(See Supplementary Note 1 for details). 
Fig.~\ref{fig5:ZALMrate}a shows the relationship between the total swapping rate $r\u{tot}$ and the rate-weighted average fidelity $F_{\mathrm{exp}}$.
The blue and red curves represent the simulated rate-fidelity trade-offs for the single-mode and 16-frequency-multiplexed cases, respectively, obtained by varying the average photon number of each EPS.
The corresponding experimental results are shown by the blue and red solid circles.
For comparison, previous demonstrations of entanglement swapping using broadband sources~\cite{Wu_2013,Jin2015,Tsujimoto2018,Samara_2021,wang2026} are shown as open circles, all of which lie below the simulated single-mode rate-fidelity trade-off for our experimental system.
The ultrafast repetition rate of the SPDC source enables this superior single-mode rate-fidelity trade-off.
The 16-channel frequency multiplexing further improves the rate-fidelity trade-off, yielding a substantially higher $r\u{tot}$ for a given $F_{\mathrm{exp}}$.

Fig.~\ref{fig5:ZALMrate}b shows the relationship between $r\u{ZALM}$ and $F\u{exp}$. 
The red dot and curve correspond to the experimental value and the simulation curve for 16 frequency multiplexing, respectively. An experimental value of $r\u{ZALM}=8.2 \times \text{10}^\text{2}$ pairs $\text{s}^{-1}$ was demonstrated with a rate-weighted average fidelity of $F\u{swap}=93.9\pm$\SI{1.4}{\%}. The green curve shows the simulated performance achievable without substantial modifications to the experimental setup, in which a BSM capable of discriminating two Bell states~($P\u{BSM}=1/2$) is implemented and the entire SPDC bandwidth of \SI{8.7}{THz} is utilized for frequency multiplexing. The experimentally measured SPDC spectrum, fitted to the measured data, is used in the simulation to account for its nonuniform spectral profile, as shown in Fig.~\ref{fig1:Concept}d.
It should be noted that, even at this stage, $r\u{ZALM}=10^5$ pairs $\text{s}^{-1}$
 can be achieved while maintaining $F\u{exp}=$\SI{91}{\%}. 
 
 As shown by the cyan curve in Fig.~\ref{fig5:ZALMrate}b, reaching the $10^9$ pairs~s$^{-1}$ regime in $r_{\mathrm{ZALM}}$ requires both denser frequency multiplexing and lower system loss. Specifically, a channel spacing of \SI{12.5}{GHz}, together with substantial reductions in system loss, is assumed, primarily by eliminating the \SI{4.8}{dB} insertion losses of the DWDM filters in the BSM node and the additional \SI{2.5}{dB} loss associated with the intrinsic heralding efficiency of \SI{57}{\%} in each signal and idler arm (see Methods for its calculation). Such a reduction in the latter loss could be achieved by using intrinsically spectrally pure photon-pair sources, such as quantum frequency-comb states, via domain-engineering~\cite{Morrison} or cavity-enhanced SPDC~\cite{Yamazaki2022,Mazeas:16,Samara:19}.
The purple curve further assumes unity spectral purity~($P=$ \SI{100}{\%}) in addition to the above improvements. Under these conditions, $r_{\mathrm{ZALM}}$ on the order of \SI{10}{MHz} remains achievable even when requiring $F\u{exp}=$\SI{99}{\%}. Compared with the theoretical proposal in Ref.~\cite{PhysRevApplied.19.054029}, the achievable $r_{\mathrm{ZALM}}$ is approximately two orders of magnitude lower for the same repetition rate and $F_{\mathrm{exp}}$. This discrepancy arises from the use of threshold detectors in the BSM. By replacing them with high-speed photon-number-resolving detectors~\cite{Cheng2023,Hao2024,Cahall:17,Endo:21}, which can effectively suppress multi-photon contributions, it would achieve both $r_{\mathrm{ZALM}}$ and $F_{\mathrm{exp}}$ comparable to those predicted in Ref.~\cite{PhysRevApplied.19.054029}.

In conclusion, we have demonstrated a ZALM Bell-pair source. Using commercially available 50-GHz DWDM filters, we implemented time-frequency-multiplexed entanglement swapping and successfully achieved the average fidelity of $93.9\pm$\SI{1.4}{\%} across the 16 ITU channel combinations. By combining frequency multiplexing with high repetition-rate operation, we realized a total swapping rate of $5.38\pm$0.17 pairs $\text{s}^{-1}$ corresponding to the ZALM Bell-pair rate of 8.2$\times\text{10}^\text{2}$ pairs $\text{s}^{-1}$. These results place our experiment in a performance regime that has not been reached by previous experiments. Moreover, our numerical simulations have identified the key challenges that must be overcome to realize $10^9$-rate ZALM Bell-pair generation. Recently, a theoretical proposal suggested exploiting BSM events involving different frequency pairs, enabling the advantage of frequency multiplexing to scale quadratically with the multiplexing factor rather than linearly~\cite{ykvm-17wh}. Together with such advances in repeater protocols, the experimental techniques demonstrated in this work represent an important step toward the practical realization of high-rate quantum repeaters.

\section*{METHODS}

\subsection*{Simulation of the spectral purity-heralding efficiency trade-off}
The spectral purity and the heralding efficiency of 
the photon pair in each DWDM channel is estimated by 
considering the joint spectral amplitude~(JSA) of the biphoton state~\cite{PhysRevLett.100.133601}. 
The biphoton state generated by SPDC is given by 
\begin {equation}
\ket{\Psi}=\iint d\omega d\omega'\Phi(\omega,\omega')\hat{a}_s^\dagger(\omega)\hat{a}_i^\dagger(\omega')\ket{\mathrm{vac}}, 
\label{eq:1}
\end{equation}
where $\omega/\omega'$ is an angular frequency of the signal/idler photon,  $\Phi(\omega,\omega')$ is the JSA of the biphoton state, and $\ket{\mathrm{vac}}$ is a vacuum state. $\hat{a}_{s/i}^\dagger(\omega)$ is the photon creation operator of the signal/idler photon whose angular frequency is $\omega$. 
We consider the case where the signal and idler photons are separated into different spatial modes, and thus the commutation relation is given by $[\hat{a}_j(\omega),\hat{a}_k^\dagger(\omega')]=\delta_{jk}\delta(\omega-\omega')$ for $j,k\in\{s,i\}$. The JSA is decomposed into  $\Phi(\omega,\omega')=\alpha(\omega,\omega')\beta(\omega,\omega')$, where $\alpha(\omega,\omega')$ and $\beta(\omega,\omega')$ are the pump envelope function and the phase matching amplitude of the nonlinear crystal, respectively. 
Finally, filter functions $F_{s/i}(\omega)$ are applied to the JSA, yielding the filtered JSA $F_s(\omega)F_i(\omega')\Phi(\omega,\omega')$.
The explicit forms of the pump envelope function $\alpha(\omega,\omega')$, the phase-matching amplitude $\beta(\omega,\omega')$, and the filter function $F_{s/i}(\omega)$ are approximated by
\begin{align}
\alpha(\omega,\omega')
&=
\operatorname{sech}\!\left(
\frac{\omega+\omega'-\omega_p}
{\Delta\sigma'_p}
\right),\nonumber\\
\beta(\omega,\omega')
&=
\exp\!\left[
-\frac12
\left(
\frac{(\omega+\omega'-\omega_p)^2}
{2(\Delta\sigma'_{\mathrm{pm}})^2}
\right)^n
\right],\nonumber\\
F_{s/i}(\omega)
&=
\exp\!\left[
-\frac12
\left(
\frac{\omega^2}
{2(\Delta\sigma'_{\mathrm f})^2}
\right)^m
\right],\nonumber
\end{align}
where $\Delta\sigma'_{\mathrm{p}}$, $\Delta\sigma'\u{pm}$, and $\Delta\sigma'_{\mathrm{f}}$ denote the width parameters corresponding to the measured FWHMs of $\Delta\sigma_p=\SI{87}{GHz}$, $\Delta\sigma_{\mathrm{pm}}=\SI{80}{GHz}$, and $\Delta\sigma_{\mathrm{f}}=\SI{45}{GHz}$, respectively. Here, $\omega_p$ is the pump angular frequency, and $n=1.4$ and $m=2.5$ are the orders of the super-Gaussian functions.
The heralding efficiency is given by the ratio of the two-fold coincidence probability to the single-detection probability, calculated as $\iint d\omega d\omega' |F_s(\omega)F_i(\omega')\Phi(\omega,\omega')|^2/\iint d\omega d\omega' |F_s(\omega)\Phi(\omega,\omega')|^2=$\SI{57}{\%}.
The spectral purity was calculated from the singular-value decomposition (Schmidt decomposition) of the filtered JSA as $P:=\sum_j \lambda_j^4=\SI{97}{\%}$, where $\lambda_j$ denotes the normalized $j$th Schmidt coefficient satisfying $\sum_j\lambda_j^2=1$.
Fig.~\ref{fig1:Concept}f was calculated by varying $\Delta\sigma_{\mathrm{pm}}$ while keeping $\Delta\sigma_p$ and $\Delta\sigma_f$ fixed.
The validity of the simulation model is independently confirmed by the good agreement between the simulated and experimentally estimated spectral purity for the case where spectral filtering is applied only to the idler photons~(see Supplementary Note 2 for details).

\subsection*{DATA AVAILABILITY}
The authors declare that the data supporting the findings of this study are available within the paper, its supplementary information files, and
Figshare.

\section*{ACKNOWLEDGEMENTS}
This work was supported by Japan Society for the Promotion of Science (JP18K13487, JP20K14393, JP22K03490), JST FOREST Program (JPMJFR222V) and R\&D of ICT Priority Technology Project (JPMI00316).

\section*{AUTHOR CONTRIBUTIONS}
Y.T. and R.I. conceived the idea, Y.T. and K.W. designed the experimental setup. 
Y.T. and D.I. conducted the experiment with the assistance of K.W., Y.T. and D.I. analyzed the data and Y.T. performed the simulation. 
R.I. and K.W. contributed to the logical development and presentation of the work. K.W. developed and constructed the electro-optic comb system. 
All authors contributed to discussions on the theory and experiment.
Y.T. wrote the manuscript with the help of all the other authors.

\section*{COMPETING INTERESTS}
The authors declare no competing interests.

\clearpage

\renewcommand{\thefigure}{S\arabic{figure}}
\setcounter{table}{0}
\renewcommand{\thetable}{S\arabic{table}}
\setcounter{figure}{0}
\renewcommand{\theequation}{S\arabic{equation}}
\setcounter{equation}{0}

\section*{Supplementary information for \enquote{Experimental zero-added-loss multiplexing Bell-pair source for long-haul quantum networks}}

\section*{SUPPLEMENTARY NOTE 1: Simulation using Gaussian-state characteristic-function method}

We present a detailed procedure for calculating the relation between the fidelity $F\u{exp}$ and the total swapping rate $r\u{tot}$ or $r\u{ZALM}$.  First, the four-fold coincidence probabilities for each measurement basis are calculated, from which $r\u{tot}$ is obtained. Next, using the probability distributions corresponding to 16 independent basis states, the density operator is reconstructed via quantum state tomography and the fidelity is evaluated. This procedure is identical to that applied to the experimentally obtained probability distributions. The coincidence probabilities are calculated based on Ref.~\cite{Takeoka_2015,PhysRevA.93.042328,Tsujimoto_2020} using the Gaussian-state characteristic-function method. Since the theoretical model and detailed procedures closely follow those in Ref.~\cite{Tsujimoto_2020}, we describe only the overall framework here. We define a density operator acting on the $N$-dimensional Hilbert space $\mathcal{H}^{\otimes N}$ as $\rho$. 
 The characteristic function of $\rho$ is defined by 
 \begin{equation}
 \chi(\bm{\xi})=\mathrm{Tr}[\rho\mathcal{W}(\bm{\xi})],
 \end{equation}
 where 
 \begin{equation}
 \mathcal{W}(\bm{\xi})
=
\exp\!\left(-i\bm{\xi}^T\bm{R}\right)
 \end{equation}
 is the Weyl operator. Here,
$\bm{R}=(x_1,\ldots,x_N,p_1,\ldots,p_N)^T$
is the vector of quadrature operators, and
$\bm{\xi}=(\xi_1,\ldots,\xi_{2N})^T\in\mathbb{R}^{2N}$
is a real-valued phase-space vector.
 When the characteristic function of the quantum state has  a Gaussian distribution as 
\begin{equation}
\chi(\bm{\xi})
=
\exp\!\left(
-\frac{1}{4}\bm{\xi}^{T}\bm{\gamma}\bm{\xi}
-i\bm{d}^{T}\bm{\xi}
\right),
\end{equation}
 the quantum state is simply characterized  by a 2$N$ $\times$ 2$N$ matrix $\bm{\gamma}$~(the covariance matrix) 
 and a 2$N$ dimensional vector $\bm{d}$~(the displacement vector). 
As shown in Supplementary Fig.~\ref{figS1:Model}a, each EPS contains a pair of two-mode squeezed vacuum states~(TMSVs), generating photon pairs in the H and V polarizations, respectively.
For example, the characteristic function of the output state from EPS~I is given by 
  \begin{equation}
 \bm{\gamma}^\mathrm{I}_{H_1V_1H_3V_3}(\mu^{\mathrm{I}}_{H},\mu^{\mathrm{I}}_{V})=\bm{\gamma}^+_{H_1V_1H_3V_3}(\mu^{\mathrm{I}}_{H},\mu^{\mathrm{I}}_{V})\oplus\bm{\gamma}^-_{H_1V_1H_3V_3}(\mu^{\mathrm{I}}_{H},\mu^{\mathrm{I}}_{V}), 
 \end{equation}
where the matrix elements are expressed as
 \begin{equation}
 \bm{\gamma}^\pm_{H_1V_1H_3V_3}(\mu^{\mathrm{I}}_{H},\mu^{\mathrm{I}}_{V})=
 	\begin{pmatrix}
 	2\mu^{\mathrm{I}}_{H}+1&0&\pm2\sqrt{\mu^{\mathrm{I}}_{H}(\mu^{\mathrm{I}}_{H}+1)}&0\\
 	0&2\mu^{\mathrm{I}}_{V}+1&0&\pm2\sqrt{\mu^{\mathrm{I}}_{V}(\mu^{\mathrm{I}}_{V}+1)}\\
 	\pm2\sqrt{\mu^{\mathrm{I}}_{H}(\mu^{\mathrm{I}}_{H}+1)}&0&2\mu^{\mathrm{I}}_{H}+1&0\\
 	0&\pm2\sqrt{\mu^{\mathrm{I}}_{V}(\mu^{\mathrm{I}}_{V}+1)}&0&2\mu^{\mathrm{I}}_{V}+1
 	\end{pmatrix}.
 	\nonumber
 \end{equation}
Here, $\mu_{p}^{k}$ denotes the average photon number of the $p$-polarized ($p=H,V$) TMSV state generated by EPS~$k$ ($k=\mathrm{I},\mathrm{I\!I}$).
Therefore, the covariance matrix of the input state
is given by 
\begin{equation}
    \bm{\gamma}\u{in}:=\bm{\gamma}^\mathrm{I}_{H_1V_1H_3V_3}(\mu^{\mathrm{I}}_{H},\mu^{\mathrm{I}}_{V})\oplus\bm{\gamma}^\mathrm{I\!I}_{H_2V_2H_4V_4}(\mu^{\mathrm{I\!I}}_{H},\mu^{\mathrm{I\!I}}_{V}). 
\end{equation}
 \begin{figure}[t]
 \begin{center}
  \includegraphics[width=\columnwidth]{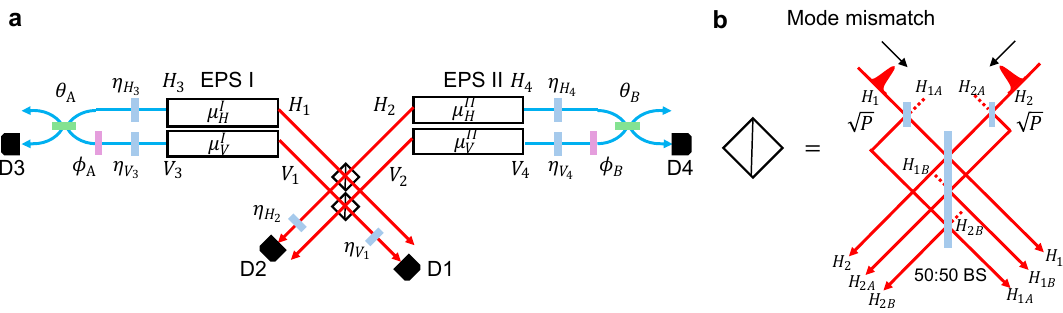}
  \caption{{\bf Theoretical model of entanglement swapping.} {\bf a} A pair of two-mode squeezed vacuum states is used to prepare
polarization entangled photon pairs. The Bell-state measurement is realized by mixing photons in modes 1 and 2 
by a 50:50 beamsplitter~(BS) followed by polarization measurements.
Polarization rotations on photons in modes 3 and 4 are realized by phase shifters and 
BSs in polarization degree of freedom. All of the photon detectors are threshold detectors.
{\bf b} Mode mismatch is modeled by virtual BSs with a transmittance of $\sqrt{P}$..
  \label{figS1:Model}} 
 \end{center}
\end{figure}
Next, the photons in modes 1 and 2 used for the BSM are mixed at a BS.
The BSM including mode mismatch is modeled following Ref.~\cite{Takeoka_2015,Tsujimoto_2020}, as illustrated in Supplementary Fig.~\ref{figS1:Model}b. 
We assume that both input photons possess the same Schmidt-mode distribution, such that the only source of imperfect interference is the mixedness arising from the finite spectral purity, while any relative spectral or temporal mismatch between the photons is neglected.
Specifically, a single input mode is split into an interfering and a non-interfering mode using a virtual BS with transmittance of $\sqrt{P}$, and the interfering modes are subsequently mixed at a 50:50 BS, where $P$ is the spectral purity defined in the main manuscript. 
Here, we introduce the ancillary vacuum modes $H_{1A},H_{1B},H_{2A},H_{2B}$ and $V_{1A},V_{1B},V_{2A},V_{2B}$ for the H and V polarizations, respectively. The characteristic function after mode mixing can then be written as
\begin{equation}
\bm{\gamma}\u{BSM}=\bm{S}^T\u{BS}\bm{S}^T\u{MM}(\bm{\gamma}\u{in}\oplus \bm{I}_{H_{1A},H_{1B},H_{2A},H_{2B}}\oplus \bm{I}_{V_{1A},V_{1B},V_{2A},V_{2B}})\bm{S}\u{MM}\bm{S}\u{BS}
\end{equation}
where
\begin{equation}
\bm{S}\u{BS}:=
\prod_{\nu \in \{H,V\}}
\left(
\bm{S}^{1/2}_{\nu_1 \nu_2}
\bm{S}^{1/2}_{\nu_{1B} \nu_{2A}}
\bm{S}^{1/2}_{\nu_{1A} \nu_{2B}}
\right)
\end{equation}
and
\begin{equation}
\bm{S}\u{MM}:=
\prod_{\nu \in \{H,V\}}
\left(
\bm{S}^{\sqrt{P}}_{\nu_1 \nu_{1A}}
\bm{S}^{\sqrt{P}}_{\nu_{2} \nu_{2A}}
\right)
\end{equation}
are the symplectic matrices corresponding to the Gaussian unitary operation for mode matching and mode mixing using a 50:50 BS, respectively. 
Here, we define the BS transformation between modes $i$ and $j$ with transmittance $t$ by $\bm{S}^t_{ij}$. 

The loss transformation is then applied to every detected mode as 
\begin{equation}
\bm{\gamma}\u{L}=\mathcal{L}(\bm{\gamma}\u{BSM})
\end{equation}
with
\begin{equation}
\mathcal{L}
=
\mathcal{L}_{V_1}^{\eta_{V_1}}\circ\mathcal{L}_{V_{1A}}^{\eta_{V_{1}}}\circ\mathcal{L}_{V_{1B}}^{\eta_{V_{1}}}\circ\mathcal{L}_{H_2}^{\eta_{H_2}}\circ\mathcal{L}_{H_{2A}}^{\eta_{H_{2}}}\circ\mathcal{L}_{H_{2B}}^{\eta_{H_{2}}}\circ\mathcal{L}_{H_3}^{\eta_{H_3}}\circ\mathcal{L}_{V_3}^{\eta_{V_3}}\circ\mathcal{L}_{H_4}^{\eta_{H_4}}\circ\mathcal{L}_{V_4}^{\eta_{V_4}},
\end{equation}
where $\mathcal{L}^{\eta}_j$ denotes the operation corresponding to the loss with transmittance $\eta$ acting on the mode $j$.

Projective measurements onto the basis states of the photons in modes 3 and 4 are simulated by applying appropriate phase shifts between the H- and V-polarization components, followed by interference at a polarization beam splitter and detection of the H-polarized component.
The covariance matrix after the polarization rotation is then given by 
\begin{equation}
\bm{\gamma}_{\mathrm{Final}}
=
\bm{M}^T
\bm{\gamma}_{\mathrm{L}}
\bm{M},
\end{equation}
where
\begin{equation}
\bm{M}
=
\bm{R}_{V_3}(\phi_A)
\bm{R}_{V_4}(\phi_B)
\bm{S}_{H_3V_3}^{\cos^2\theta_B}
\bm{S}_{H_4V_4}^{\cos^2\theta_A}.
\end{equation}
Here, $\bm{R}_j(\phi)$ is the symplectic matrix corresponding to the phase shift $\phi$ on mode $j$. 
In the entanglement swapping experiment, we choose the event where H and V polarized photons are respectively detected by D1 and D2 as a successful event as show in Supplementary Fig.~\ref{figS1:Model}. Note that the maximum number of successful events allowed for a BSM using linear optical elements is four times the value described above.

Finally, the four-fold coincidence probability is given by 
\begin{align}
P\u{\mathrm{4fold}}(\phi_A,\theta_A,\phi_B,\theta_B)
&= \mathrm{Tr}\!\left[
\rho^{\gamma\u{Final}} \prod_{i=1}^{4} \Pi\u{on}^{B_i}
\right] \\
&= \sum_{S \subseteq \{1,2,3,4\}}
(-1)^{|S|}
\frac{2^{|B_S|}}{\sqrt{\det\!\bigl(\bm{\gamma}_{B_S} + I\bigr)}}, 
\label{eq:4fold}
\end{align}
where
\begin{equation}
\Pi\u{on}^{B_i} = I-\ketbra{0}{0}_{B_i}
\end{equation}
is the POVM element of the photon detection in mode $B_i$ and 
\begin{align}
B_1 &= \{H_2, H_{2A}, H_{2B}\}, \nonumber\\
B_2 &= \{V_1, V_{1A}, V_{1B}\}, \nonumber\\
B_3 &= \{H_3\}, \nonumber\\
B_4 &= \{H_4\}, \nonumber\\
B_S &= \bigcup_{i \in S} B_i. \nonumber
\end{align}
andwhere $\bm{\gamma}_{B_S}$ denotes the principal submatrix of $\bm{\gamma}$ corresponding to the modes in $B_S$, with $\bm{\gamma}_{B_\varnothing}$ understood as the $0\times0$ matrix.
In Supplementary Eq.~(\ref{eq:4fold}), we have used the identity for the overlap between an $m$-mode Gaussian state with covariance matrix $\bm{\gamma}$ and the $m$-mode vacuum state as
\begin{equation}
\mathrm{Tr}\left[\rho^\gamma\ketbra{0}{0}^{\otimes m}\right]=\frac{2^m}{\sqrt{\mathrm{det}(\bm{\gamma}+\bm{I})}}.
\end{equation}
The four-fold coincidence probability
$P_{\mathrm{4fold}}(\phi_A,\theta_A,\phi_B,\theta_B)$
is calculated using the experimentally characterized system parameters summarized in Supplementary Table~\ref{tableS1:Parameters}, including the Klyshko efficiencies and the spectral purity~$P$, together with the average photon number of each TMSV state. The quantum efficiencies of the detectors and the heralding efficiencies are included in the Klyshko efficiency.
It is important to note that $P_{\mathrm{4fold}}$ is not equivalent to the success probability of the local BSM. The latter includes erroneous events in which two photon pairs are generated from one EPS while no photon pair is generated from the other. Such events do not contribute to the generation of swapped Bell pairs and are therefore excluded from $r_{\mathrm{ZALM}}$.

\begin{table}[h]
\caption{{\bf The experimental parameters used for the simulation}}\label{tableS1:Parameters}%
\begin{tabular}{@{}llllllllll@{}}
\toprule
$\eta_{V_1}$ &  $\eta_{H_2}$ & $\eta_{H_3}$& $\eta_{V_3}$& $\eta_{H_4}$& $\eta_{V_4}$&   $P$ \\
\midrule    
\SI{7.2}{\%} & \SI{5.9}{\%} & \SI{4.2}{\%}& \SI{4.2}{\%} & \SI{4.6}{\%} & \SI{5.0}{\%} & \SI{97}{\%} \\
\botrule
\end{tabular}
\end{table}

The entanglement fidelity $F_{\mathrm{exp}}$ is evaluated from $P\u{\mathrm{4fold}}(\phi_A,\theta_A,\phi_B,\theta_B)$ for the 16 polarization projection settings corresponding to the basis states shown in Fig.~4d.

The total swapping rate $r\u{tot}$ shown in Fig.~5a is evaluated by 
\begin{equation}
r\u{tot}
=
f_{\mathrm{rep}}\,M_{\mathrm{eff}}
\sum_{\theta_A,\theta_B\in\{0,\pi/2\}}
P_{\mathrm{4fold}}(0,\theta_A,0,\theta_B),
\label{eqs17}
\end{equation}
where the average photon numbers of all TMSV states are assumed to be identical, i.e. 
$
\mu_H^{\mathrm{I}}
=
\mu_V^{\mathrm{I}}
=
\mu_H^{\mathrm{I\!I}}
=
\mu_V^{\mathrm{I\!I}}
\equiv
\mu$, and $r\u{tot}$ is evaluated as a function of the common average photon number $\mu$. Here, $f_{\mathrm{rep}}=\SI{3.0}{GHz}$ is the pump repetition rate and $M_{\mathrm{eff}}$ is the effective multiplexing factor. Specifically, no multiplexing correction is applied in the single-mode case, and thus $M_{\mathrm{eff}}=1$. For the multiplexed cases, $M_{\mathrm{eff}}$ is obtained by multiplying the nominal multiplexing number by a correction factor determined from the measured spectral profile shown in Fig.~1d, yielding $M_{\mathrm{eff}}=0.95M$ for the \SI{800}{GHz} bandwidth and $M_{\mathrm{eff}}=0.86M$ for the full \SI{8.7}{THz} SPDC bandwidth, where $M$ denotes the nominal multiplexing number.

The ZALM Bell-pair rate $r_{\mathrm{ZALM}}$ shown in Fig.~5b is evaluated by removing the detection losses in modes~3 and~4 while retaining the intrinsic heralding efficiency associated with spectral filtering. Accordingly, the system transmittances are set to $\eta_{H_3}=\eta_{V_3}=\eta_{H_4}=\eta_{V_4}=\SI{57}{\%}$, where \SI{57}{\%} is the intrinsic heralding efficiency due to spectral filtering~(see Methods for details). When assuming an ideal linear-optical BSM with a success probability of $P_{\mathrm{BSM}}=1/2$, the four-fold coincidence probability is multiplied by a factor of four. Furthermore, for the idealized scenarios represented by the cyan and purple curves, perfect transmission is assumed for all modes, i.e.,
$\eta_{V_1}=\eta_{H_2}=\eta_{H_3}=\eta_{V_3}=\eta_{H_4}=\eta_{V_4}=1$.

\section*{Supplementary Note 2: Experimental validation of the simulated spectral purity}
To validate the numerical model used to estimate the spectral purity in Methods, we compare the simulated spectral purity with that experimentally inferred from measurements of the intensity correlation function of the idler photons without detecting the signal photons. Since the source is pulsed, we denote by $g_n^{(2)}$ the intensity correlation between $n$th neighbor pulses, where $g_0^{(2)}$ corresponds to the autocorrelation function. The relationship between $g_0^{(2)}$ and $P$ can be derived as follows. We consider a statistical mixture of light in $K$ independent Schmidt modes. Let $\lambda_i$ denote the normalized Schmidt coefficient of the $i$th Schmidt mode, satisfying $\sum_{i=1}^{K}\lambda_i^2=1$. In the low-gain regime, the average photon number in the $i$th Schmidt mode is proportional to $\lambda_i^2$, and we define the normalized modal weight as
\begin{equation}
p_i=\frac{\langle N_i\rangle}{\sum_j\langle N_j\rangle}\simeq\lambda_i^2,
\end{equation}
where $N_i$ is the photon-number operator for the $i$th Schmidt mode.

The intensity correlation within a single pulse is then given by
\begin{eqnarray}
g_0^{(2)}
&=&
\frac{\left\langle:\left(\sum_{i=1}^{K}N_i\right)^2:\right\rangle}
{\left\langle\sum_{i=1}^{K}N_i\right\rangle^2}
\nonumber\\
&=&
2\sum_{i=1}^{K}p_i^2
+
2\sum_{i>j}p_ip_j,
\end{eqnarray}
where $\langle:\cdots:\rangle$ denotes the normally ordered expectation value. The second equality follows from the statistical independence of the Schmidt modes and the fact that each reduced Schmidt mode of the SPDC state exhibits an ideal thermal photon-number distribution. Using
\begin{equation}
2\sum_{i>j}p_ip_j
=
1-\sum_ip_i^2,
\end{equation}
we obtain
\begin{equation}
g_0^{(2)}
=
1+\sum_ip_i^2
\simeq
1+\sum_i\lambda_i^4
=
1+P,
\end{equation}
where the last equality follows from the definition of the spectral purity,
\begin{equation}
P=\sum_i\lambda_i^4.
\end{equation}
This result is consistent with Ref.~\cite{Christ_2011}.

In the experiment of $g_0^{(2)}$ measurement, the idler photons are first passed through a DWDM filter and then split by a 50:50 FBS. 
The intensity correlation function $g_n^{(2)}$ is estimated from the measured coincidence counts as 
\begin{equation}
g_n^{(2)}=\frac{C_{12}^{(n)}f_{\mathrm{rep}}}{S_1S_2},
\end{equation}
where $C_{12}^{(n)}$ is the coincidence count between a photon detected at D1 and a photon detected at D2 separated by $n$ pulse intervals, and $S_1$ and $S_2$ are the single-count rates at D1 and D2, respectively. 
Assuming $g_{\infty}^{(2)}=1$, we obtain
$g_0^{(2)} = \frac{C_{12}^{(0)}}{C_{12}^{(\infty)}}$. Since the pulse duration is much shorter than the pulse interval, we estimated the uncorrelated coincidence count $C_{12}^{(\infty)}$ by averaging the coincidence counts from neighboring pulses, 
\begin{equation}
C_{12}^{(\infty)} = \frac{1}{4}\sum_{n=-2,-1,1,2}C_{12}^{(n)}.
\end{equation}
Using this method, we obtain $g^{(2)}_0 = 1.84\pm0.03$, corresponding to a spectral purity of $P = 84\pm\SI{3}{\%}$. This value 
agrees well with the simulated value of \SI{86}{\%}, calculated from the JSA $F_i(\omega')\Phi(\omega,\omega')$, thereby validating the numerical model used to estimate the spectral purity.


\end{document}